\documentclass[
    ,final            
  ]
  {aipproc}

\layoutstyle{8x11double}

\usepackage{amssymb,amsmath,amstext,amsthm,amsfonts}

\newcommand{\plotscale}{.65}

\newcommand{\reffig}[1]{Fig.~\ref{#1}}

\newcommand{\refcite}[1]{Ref.~\cite{#1}}

\newcommand{\refetal}[1]{\emph{et~al.}~\cite{#1}}
\newcommand{\atom}[2]{\mbox{$^{#1}\text{#2}$}}
\newcommand{\carbon}{{\atom{12}{C}}}
\newcommand{\oxygen}{{\atom{16}{O}}}
\newcommand{\GeV}{{\;\mathrm{GeV}}}

\begin{document}

\title{Pion production in neutrino interactions with nuclei}



\author{T.~Leitner}{address={Institut f\"ur Theoretische Physik, Universit\"at Giessen, Germany}}
\author{O.~Lalakulich}{address={Institut f\"ur Theoretische Physik, Universit\"at Giessen, Germany}}
\author{O.~Buss}{address={Institut f\"ur Theoretische Physik, Universit\"at Giessen, Germany}}
\author{U.~Mosel}{address={Institut f\"ur Theoretische Physik, Universit\"at Giessen, Germany}}
\author{L.~Alvarez-Ruso}{address={Departamento de F\'isica, Centro de F\'isica Computacional, Universidade de Coimbra, Portugal}}

\begin{abstract}
  Neutrino-induced pion production on nuclear targets is the major inelastic channel in
  all present-day neutrino-oscillation experiments. It has to be understood quantitatively
  in order to be able to reconstruct the neutrino-energy at experiments such as MiniBooNE
  or K2K and T2K. We report here results of cross section calculations for both this
  channel and for quasielastic scattering within the semiclassical GiBUU method. This
  methods contains scattering, both elastic and inelastic, absorption and side-feeding of
  channels all in a unitary, common theoretical framework and code.  We find that charged
  current quasielastic scattering (CCQE) and $1\,\pi$ production are closely entangled in
  actual experiments, due to final state interactions of the scattered nucleons on one
  hand and of the $\Delta$ resonances and pions, on the other hand. We discuss the
  uncertainties in the elementary pion production cross sections from ANL and BNL.  We
  find the surprising result that the recent $1 \pi$ production cross section data from
  MiniBooNE are well described by calculations without any FSI. For higher energies we
  study the validity of the Bloom-Gilman quark-hadron duality for both electron- and
  neutrino-induced reactions. While this duality holds quite well for nucleon targets, for
  nuclear targets the average resonance contributions to the structure function $F_2$ are
  always lower than the DIS values.  This result indicates a significant impact of nuclear
  effects on observables, reducing the cross section and structure functions by at least
  30-40$\%$ and changing the form of various distributions.
\end{abstract}

\maketitle


\section{Introduction}

Neutrino oscillations experiments search for a distortion in the
neutrino flux at the detector positioned far away from the source.  By comparing both
near and far neutrino energy spectra, one gains information about the oscillation
probability and with that about mixing angles and mass squared differences.  However, the
neutrino energy, that enters critically into the oscillation probability, is not directly measurable
but has to be reconstructed from the reaction
products. A proper understanding of neutrino-nucleus interactions is, therefore, essential for the
interpretation of current neutrino oscillations experiments.  Present $\nu_\mu$ disappearance experiments use the CCQE reaction both as signal event and
to reconstruct the neutrino energy from the outgoing muon. CCQE is defined as $\nu_\ell n
\to \ell^- p$ on a single nucleon; in the nucleus, CCQE is masked by final-state
interactions (FSI). Thus, the correct identification of CCQE events is directly related to
the question of how FSI influence the event selection. The main background to CCQE is
CC$\,1\pi^+$ production. If the pion is absorbed in the nucleus and/or not seen in the
detector, these events can be misidentified as CCQE. Consequently, a proper understanding
of CCQE \emph{and} CC$\,1\pi^+$ is essential for the reconstruction of the neutrino energy.

The main task in a $\nu_e$ appearance experiment like MiniBooNE is to detect electron
neutrinos in a (almost) pure $\nu_\mu$ beam. The signal event, the $\nu_e$ CCQE
interaction, is dominated by background. A major problem comes from misidentified events,
mainly because of the fact, that the MiniBooNE detector cannot distinguish between a
photon and an electron.  Thus, $\nu_\mu$ induced neutral current $\pi^0$ production, where
the $\pi^0$ decays into two $\gamma$s, is the major source of background when one of the
photons is not seen or both Cherenkov rings overlap.

As all of the present oscillations experiments use nuclear targets, it is mandatory to
consider FSI, i.e., pion rescattering, with and without charge exchange, and absorption in
the nuclear medium. A realistic treatment of the final state interactions (FSI) can be achieved in the framework of a
coupled-channel transport theory --- the GiBUU model.

After a brief review of our model, we first discuss the impact of pion production on CCQE
measurements. We further investigate the influence of nuclear effects on CC$1\pi^+$ and
NC$1\pi^0$ cross sections, and, where possible, we confront our model to recent data
measured at MiniBooNE.  In the second part we extend the description to higher energies
using duality arguments.

\section{GiBUU model}

We treat neutrino-nucleus scattering as a two-step process. In the initial-state step, the
neutrinos interact with nucleons embedded in the nuclear medium.  In the final-state step,
the outgoing particles of the initial reaction are propagated through the nucleus using a
hadronic transport approach.

In the energy region relevant for MiniBooNE, SciBooNE and K2K, the elementary $\nu N$
reaction is dominated by two processes: quasielastic scattering and the excitation of the
$\Delta$ resonance (P$_{33}$(1232)). Additionally, our model includes the excitation of 12
$N^*$ and $\Delta$ resonances with invariant masses less than 2 GeV and also a
non-resonant single-pion background. Details are given in \cite{Leitner:2008ue}.

The excitation of the resonances (R) is described within the isobar model with the help of the
nucleon, $N-\Delta$ and $N-N^*$ form factors.  Vector form factors are derived from MAID
helicity amplitudes \cite{MAIDWebsite,Tiator:2006dq,Drechsel:2007if} extracted from
electron scattering experiments.

\begin{figure}[tb]
\includegraphics[width=\columnwidth]{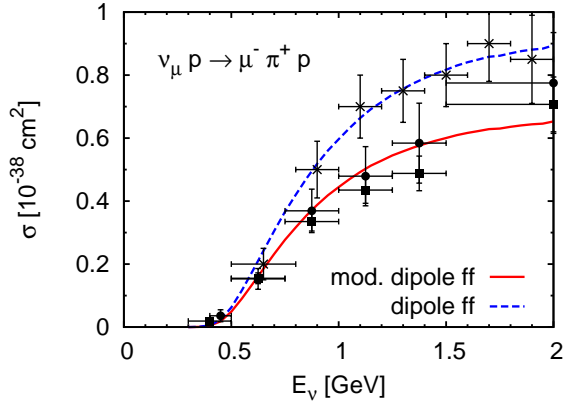}
\caption{Total $\nu_\mu p \to \mu^- \pi^+ p$ cross section as a function of
  the neutrino energy compared to the pion production data of of ANL
  (Refs.~\cite{Barish:1978pj} ($\bullet$), \cite{Radecky:1981fn} ($\blacksquare$)) and BNL
  (\cite{Kitagaki:1986ct} ($\times$)). The solid line has been obtained with a form factor
  fitted to the ANL data, the dashed one is fitted to the BNL data.}
    \label{fig:Delta_sigmatot}
\end{figure}

Experimental information on the N-R axial form factors is very limited.
Goldberger-Treiman relations have been derived for the axial couplings, but there is no
information about the $Q^2$ dependence. We apply PCAC and pion pole dominance to derive
one of the axial couplings for each resonance and to relate it to the pseudoscalar
coupling.  The $Q^2$ dependence of the $\Delta$ axial form factors are fitted to either
ANL or BNL bubble chamber neutrino-scattering $d\sigma/dQ^2$ data for the $\nu_\mu p \to
\mu^- \pi^+ p$ reaction. Fig.~\ref{fig:Delta_sigmatot} shows the integrated cross section
together with the data.  We note already here, that the solid curve fits the ANL data,
while the dashed curve fits the BNL data. Thus, the latter would obviously lead to higher
pion production cross section also on the nucleus.

\begin{figure}[tb]
  \centerline{\includegraphics[width=\columnwidth]{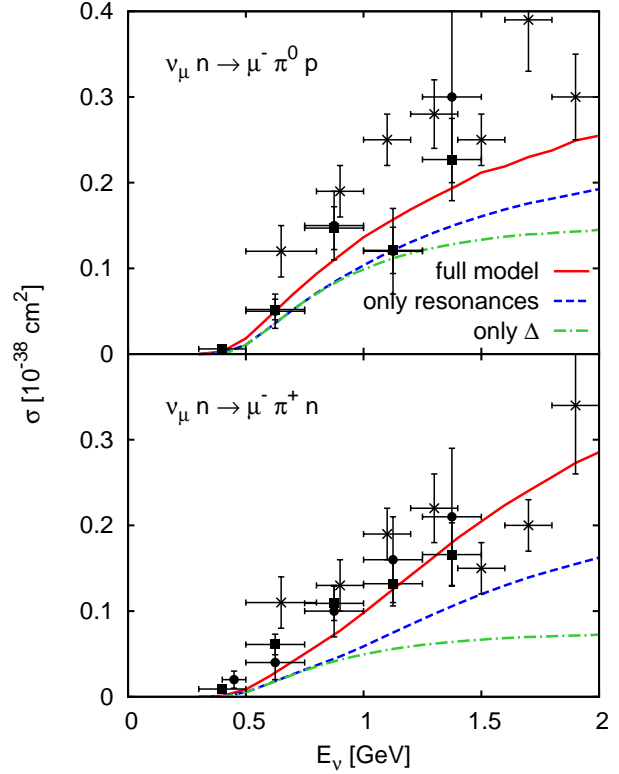}}
  \caption{Total CC pion production cross sections for the mixed isospin
    channels as a function of the neutrino energy compared to the pion production data of
    of ANL (Refs.~\cite{Barish:1978pj} ($\bullet$) and \cite{Radecky:1981fn}
    ($\blacksquare$)) and BNL (\cite{Kitagaki:1986ct} ($\times$)). The solid lines denote
    the our full result including the non-resonant background.  Furthermore, we show the
    results for pion production only through the excitation and the subsequent decay of
    all resonances (dashed lines) or through the $\Delta$ alone (dash-dotted lines). Note
    that the ANL based fit for the $\Delta$ axial form factor has been used here. }
    \label{fig:xsec_pionprod_wBG}
\end{figure}

The single-$\pi$ non-resonant background cross section $\sigma_\text{BG}$ includes vector,
axial and also interference contributions
\begin{equation}
  \sigma_{BG}  =  \sigma_\text{BG}^{\text{V}} + \sigma_\text{BG}^{\text{A}}+ \sigma_\text{BG}^{\text{V/A}}
  =  \sigma_\text{BG}^{\text{V}} +  \sigma_\text{BG}^{\text{non-V}} \ .
\end{equation}
The vector part is fully determined by electron scattering data, as described in
\cite{Leitner:2008ue}.  The axial and the interference term collected under the label
``non-V'' are only present in neutrino scattering and are fitted to the available neutrino
data for both, $\nu_\mu n \to \mu^- \pi^{+} n$ and $\nu_\mu n \to \mu^- \pi^{0} p$. The
final results are shown in Fig.~\ref{fig:xsec_pionprod_wBG}.

\begin{figure}[tb]
  \includegraphics[scale=\plotscale]{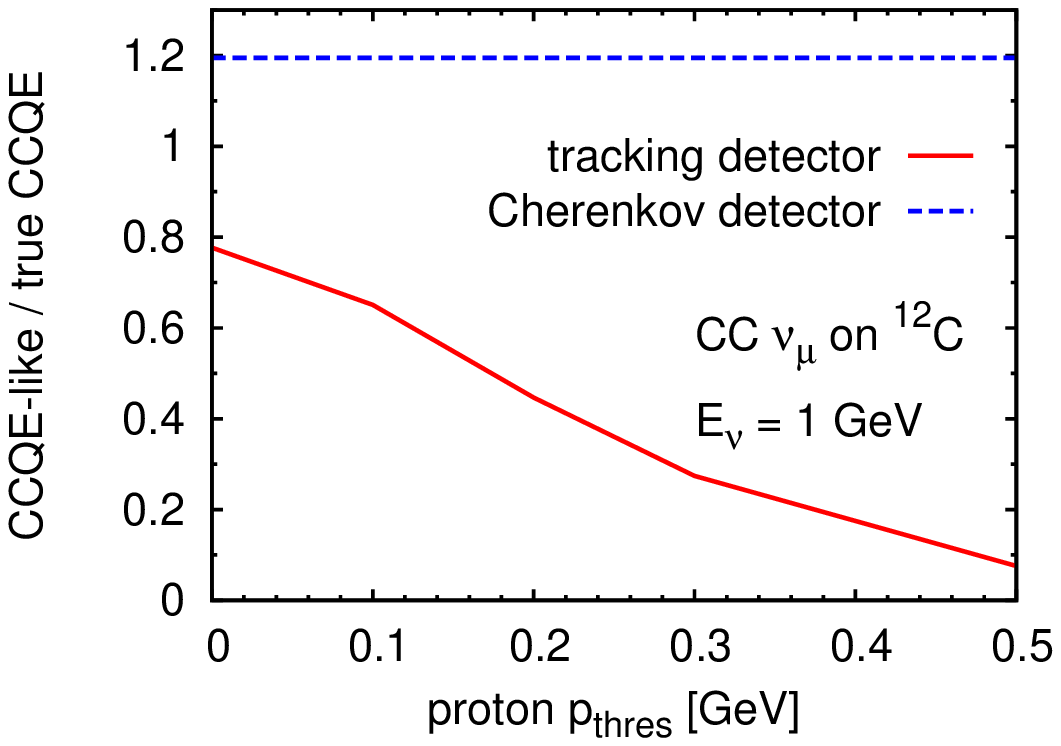}
  \includegraphics[scale=\plotscale]{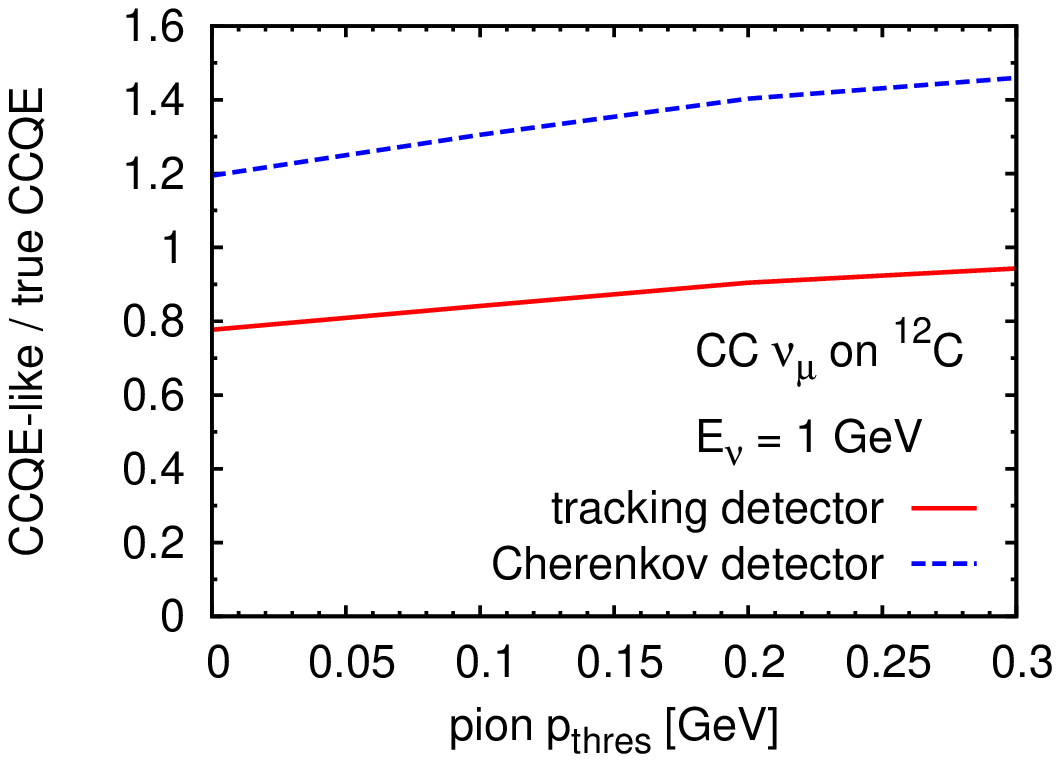}
    \caption{Ratio of the CCQE-like to the true CCQE cross section as a function of
  the proton (pion) momentum detection threshold for CC $\nu_\mu$ on \carbon{} at
  $E_\nu=1$ GeV. The solid lines are obtained using the tracking detector identification,
  while the dashed lines are for Cherenkov detectors.
  \label{fig:CCQElike_over_trueCCQE_thresholds}}
\end{figure}

The neutrino-nucleon cross sections are modified in the nuclear medium. Bound nucleons are
treated within a local Thomas-Fermi approximation. They are exposed to a mean-field
potential which depends on density and momentum. We account for this by evaluating the
above cross sections with full in-medium kinematics, i.e., hadronic tensor and phase-space
factors are evaluated with in-medium four-vectors. We also take Pauli blocking and
collisional broadening of the outgoing hadrons into account. Our model for
neutrino-(bound)nucleon scattering is described in detail in \cite{Leitner:2008ue}.

\begin{figure}[tb]
  \includegraphics[scale=\plotscale]{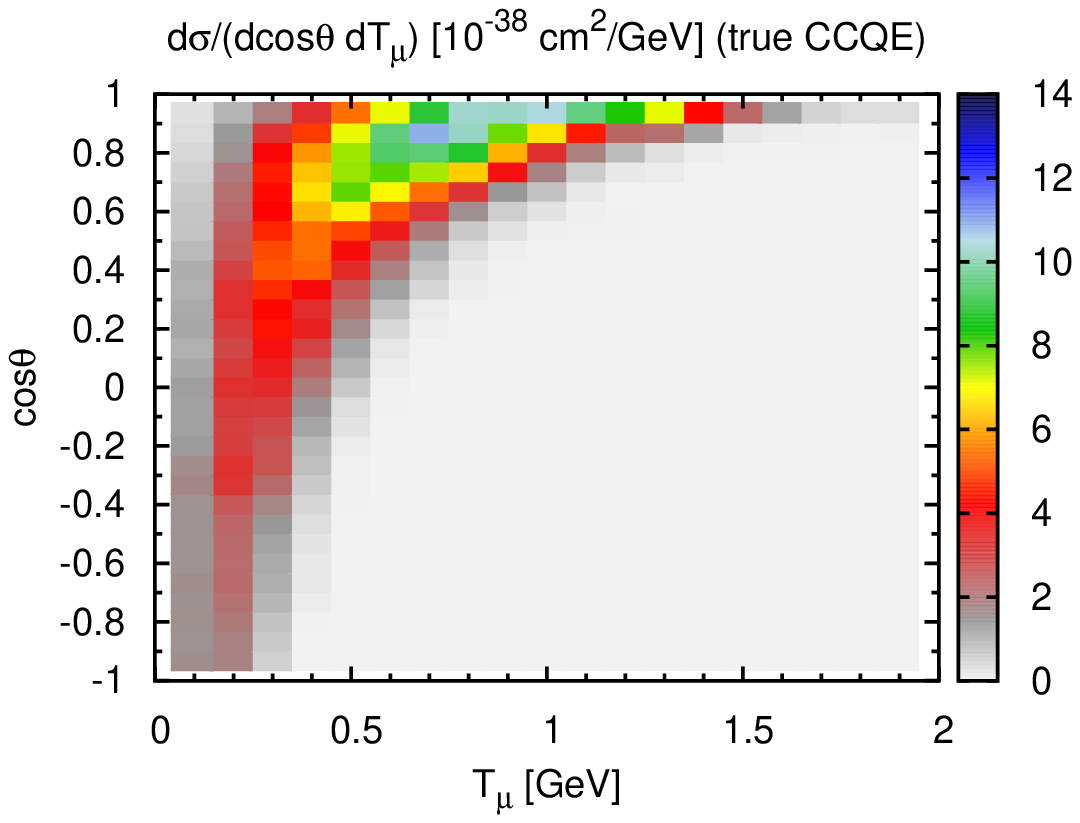}
  \includegraphics[scale=\plotscale]{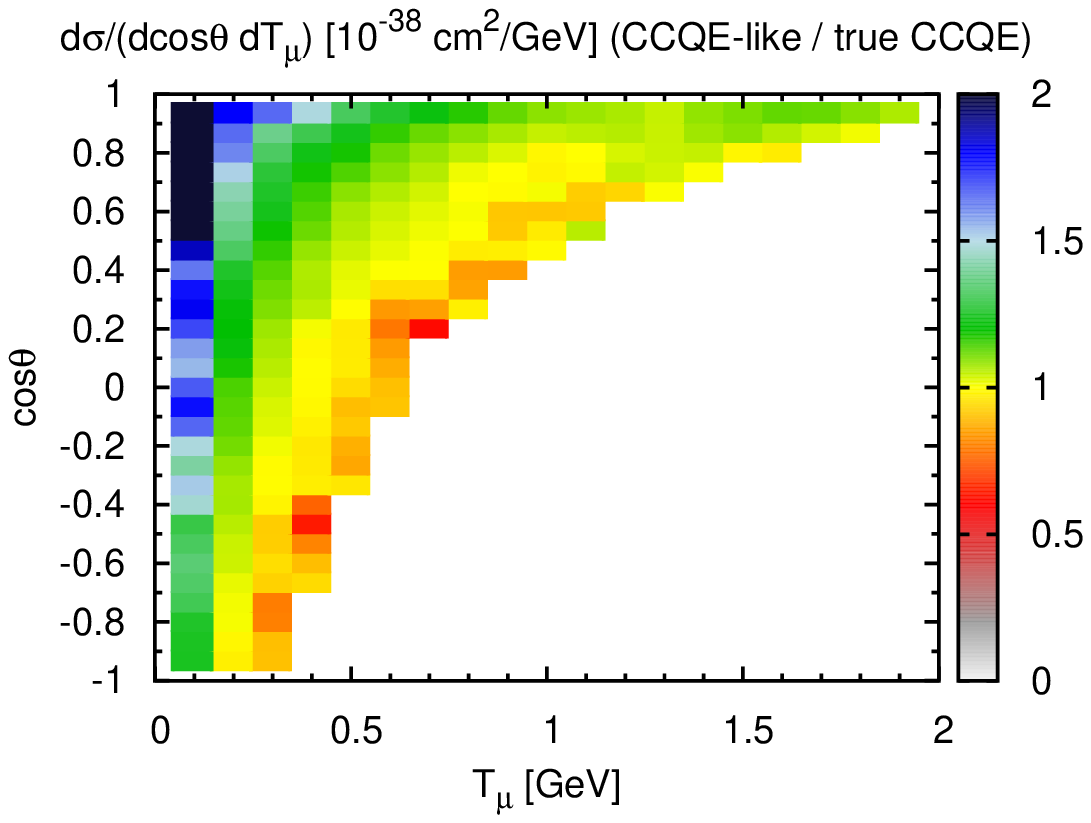}
    \caption{Double differential cross section on \carbon{} averaged over the
  MiniBooNE flux as a function of the muon kinetic energy and the muon scattering angle.
  The left panel shows the true CCQE cross section, the right panel the ratio of the CCQE-like to the true CCQE cross section.
  \label{fig:MiniBooNE_doublediff_QE}}
\end{figure}

After the initial neutrino-nucleon interaction, the produced particles propagate through and out of
the nucleus. During propagation they undergo FSI which are simulated with the
coupled-channel semi-classical GiBUU transport model (for details, see \refcite{gibuu} and
references given there).  It is based on the BUU equation which describes the space-time
evolution of a many-particle system in a mean-field potential including a collision term.
Nucleons and resonances acquire medium-modified spectral functions and are propagated
off-shell. Herby we ensure, that vacuum spectral functions are recovered after leaving the
nucleus.  The collision term accounts for changes (gain and loss) in the phase-space
density due to elastic and inelastic collisions between the particles, and also to
particle decays into other hadrons. Baryon-meson two-body interactions (e.g., $\pi N \to
\pi N$) are described by resonance contributions and a small non-resonant background term;
baryon-baryon cross sections (e.g., $NN \to NN$, $R N \to N N$, $R N \to R' N$, $N N \to
\pi NN$) are either fitted to data or calculated, e.g., in pion exchange models. The
three-body channels $\pi N N \to NN$ and $\Delta N N \to NNN$ are also included. The BUU
equations for all particle species are thus coupled through the collision term and also
through the potentials. Such a coupled-channel treatment is required to account for side
feeding into different channels.

The treatment of pion final state interactions in GiBUU has been widely tested both with $\pi A$ and
$\gamma A \to \pi X $ data \cite{Krusche:2004uw,Buss:2006vh,Buss:2006yk}. The latter reaction is quite similar to
the $\nu A \to \pi X$ reaction in that the incoming particle interacts with all target nucleons and the vector couplings are the same.

\section{CCQE and CC$1\pi^+$ entanglement}

One challenge is to identify \emph{true} CCQE events in the detector, i.e., muons
originating from an initial QE process.  The difficulty comes from the fact that the true CCQE events
are masked by FSI in a detector built from nuclei. In general, at Cherenkov detectors such as MiniBooNE CCQE-like events are all those where
no pion is detected while in tracking detectors such as K2K-SciBar/SciFi CCQE-like events
are those where a single proton track is visible and at the same time no pions are
detected. In both detector types the FSI lead to misidentified events,
e.g., an initial $\Delta$ whose decay pion is absorbed or which undergoes ``pion-less
decay'' can count as CCQE event --- we call this type of background events ``fake CCQE''
events. We denote every event which looks like a CCQE event by ``CCQE-like''.

To investigate the relation between the CCQE-like and true CCQE cross section, we show
their ratio as a function of proton and pion momentum thresholds in
\reffig{fig:CCQElike_over_trueCCQE_thresholds}. As the proton is not at all relevant for
the CCQE identification in Cherenkov detectors, the ratio is independent of the proton
momentum detection threshold (dashed line in left panel). This is very different in
tracking detectors which rely on the detected proton --- here the efficiency is reduced to
$\approx$10\% at a proton momentum threshold of 0.5 GeV (solid line in left panel). Even
at $|\vec{p}|_\text{thres}^p=0$ the efficiency does not exceed 80\% because of
charge-exchange processes that lead to the emission of undetected neutrons and because of
secondary proton knockout that leads to multiple-proton tracks.  Focussing on the right
panel of \reffig{fig:CCQElike_over_trueCCQE_thresholds} we find that the CCQE-like cross
section increases for both detector types as $|\vec{p}|_\text{thres}^\pi$ increases. In
this case even more events with pions in the final state appear as CCQE-like because these
pions are below threshold and thus not detected.

Fixing the flux normalization with HARP's pion-production data, the MiniBooNE
collaboration has presented their first, preliminary absolutely normalized total,
differential and double differential cross sections for CCQE and finds
an excess of about 35\% compared to the total cross section measured at NOMAD, ANL and BNL
\cite{Katori:2009du}.  We emphasize that these absolute cross sections
depend directly on the pion background subtraction which again is based on the Monte Carlo
prediction (cf. \refcite{Katori:2009du}).

In \reffig{fig:MiniBooNE_doublediff_QE}, we show our prediction for the double
differential cross section at MiniBooNE in muon observables, all calculated with $M_A=1$
GeV. The left panel shows the true CCQE events. To emphasize the role of ``fake'' CCQE
events, we show the ratio CCQE-like/true CCQE in the right panel. This ratio shows that
fakes contribute mainly at high energy transfers (low $T_\mu$) and forward angles. Unlike for monochromatic
beams, the QE and $\Delta$ peaks are not distinguishable any more but strongly overlap.
This fact makes a model-independent cut based on muon variables to subtract the background
impossible.

\section{MiniBooNE's CC\,1$\pi^+$ measurement}
\begin{figure}[tb]
  \centering
  \includegraphics[width=\columnwidth]{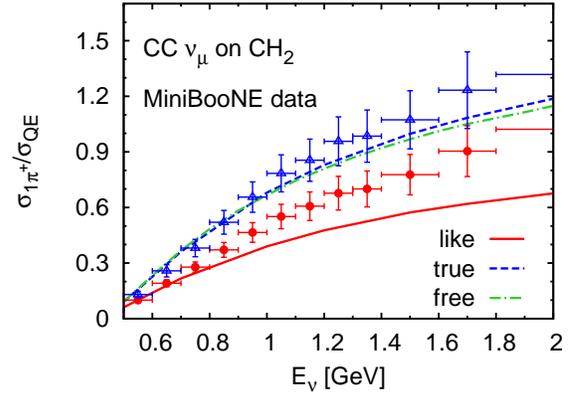}
  \caption{Single-$\pi^+$/QE cross section ratio for CC interactions vs.~neutrino energy
    on CH$_2$ together with recent data from MiniBooNE \cite{AguilarArevalo:2009eb} (upper
    data set: corrected for FSI, lower data set: uncorrected for FSI). The solid lines
    denote the CC$1\pi^+$-like/CCQE-like result (Cherenkov detector definitions), the
    dashed lines stand for the true CC$1\pi^+$/true CCQE result, and the dash-dotted lines
    give the vacuum expectation, i.e., the sum of the nucleon cross sections (with two
    additional protons in the MiniBooNE case).
    \label{fig:MB_CC}}
\end{figure}

In \reffig{fig:MB_CC} we give our results for the single-$\pi^+$/QE ratio for CC
interactions on mineral oil CH$_2$. The solid lines denote the CC$1\pi^+$-like/CCQE-like
result, the dashed lines stand for the true CC$1\pi^+$/true CCQE result, and the
dash-dotted lines give the vacuum expectation. Note that we have applied the Cherenkov
detector identification criteria.

We emphasize that nuclear corrections cancel out in the ratio, only as long as FSI are not
considered (``true'' vs.~``free''). In general, the complexity of FSI prevent such
cancellations as one can infer from the result denoted with ``like'' which does not
coincide with the ``true'' and ``free'' ones.

\begin{figure}[tb]
  \includegraphics[scale=\plotscale]{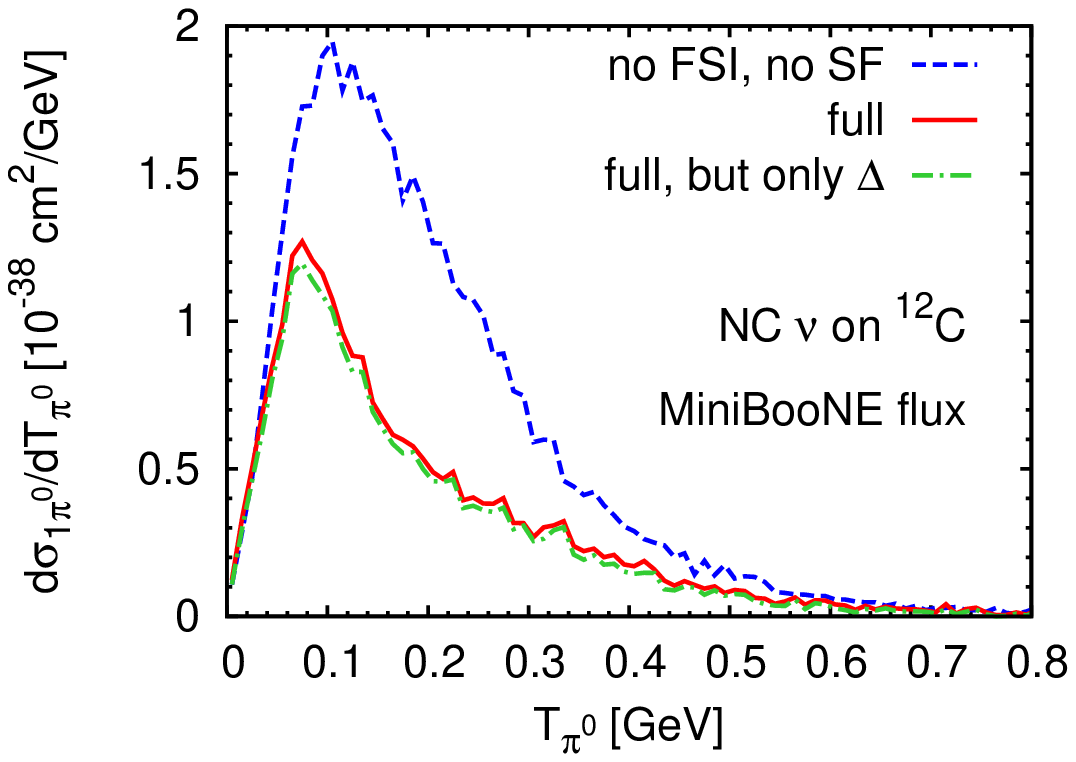}
  \includegraphics[scale=\plotscale]{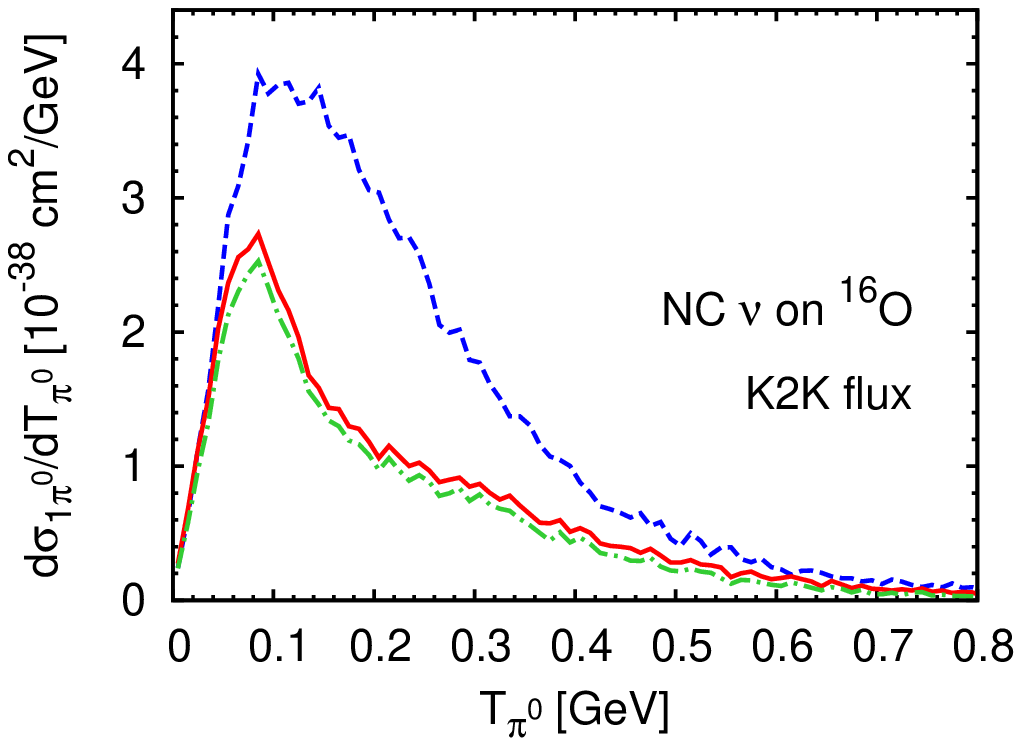}
  \caption{Left panel: NC induced single-$\pi^0$ production on \carbon{} as a function of
    the pion kinetic energy averaged over the MiniBooNE flux.
    Right panel: same on \oxygen{} averaged over the K2K
    flux. The dashed lines show our calculation without FSI or spectral functions, both
    included in the full calculation denoted with the solid lines. The dash-dotted lines
    indicate the contribution from the $\Delta$ resonance to the full calculation.
  \label{fig:MiniBooNE_K2K_NC}}
\end{figure}

We further compare to very recent MiniBooNE data \cite{AguilarArevalo:2009eb}
(\reffig{fig:MB_CC}). Let us first focus on the data denoted with the triangles (upper
data set). These are corrected for FSI using a specific Monte Carlo generator, i.e., they
give the cross sections for bound nucleons ``before FSI''. As this procedure introduces a
model dependence in the data, a fully consistent comparison is not possible.  Ignoring
this inconsistency, our calculation denoted with ``true'' should be the one to compare
with.  The agreement is perfect for energies up to 1.5 GeV, and still within their error
bars for higher $E_\nu$. The MiniBooNE data denoted with bullets (lower data set) is their
result for the ratio of CC$1\pi^+$-like to CCQE-like. As these data are not corrected for
FSI within a specific Monte Carlo event generator, this observable is less model
dependent. Still, the energy reconstruction requires specific assumptions as well as the
detector simulation.  We find that our calculation clearly underestimates the uncorrected
data. However, the perfect agreement with their corrected distribution indicates a
significant difference between the pion absorption models.

The underestimate of the pion/quasielastic ratio in particular at higher energies could be
due to, among other possibilities, an underestimate of the pion production cross section
or an overestimate of the CCQE-like cross section. Both depend directly on the input at
the nucleon level, i.e., in particular on the axial form factor $C_5^A$ of the $\Delta$
resonance (see \reffig{fig:Delta_sigmatot}).  For the results presented here, we have used
the ANL data as a reference. However, we have shown in \refcite{Leitner:2009ec} that an
increase of the total pion production cross section on the nucleon compatible with the BNL
data also seems to be insufficient to describe this ratio at all energies. We note that a
similar result for the ratio has been recently obtained by Athar \refetal{Athar:2009rc}.

\section{NC$1\pi^0$}

In the left panel of \reffig{fig:MiniBooNE_K2K_NC}, we show our results for NC
single-$\pi^0$ production off \carbon{} as a function of the pion kinetic energy.  We have
averaged over the MiniBooNE energy flux which peaks at about 0.7 GeV neutrino energy
\cite{AguilarArevalo:2008yp}.  In NC reactions the total pion yield is dominated by
$\pi^0$ production, while $\pi^+$ dominate in CC processes (for details, see
\refcite{Leitner:2008wx} and references therein).  Comparing the dashed with the solid
line (results without FSI and spectral function vs.~full calculation), one finds a
considerable change. The shape is caused by the energy dependence of the pion absorption
and rescattering cross sections.  Pions are mainly absorbed via the $\Delta$ resonance,
i.e, through $\pi N \to \Delta$ followed by $\Delta N \to NN$. This explains the reduction
in the region around $T_\pi=0.1-0.3$ GeV.  Pion elastic scattering $\pi N \to \pi N$
reshuffles the pions to lower momenta and leads also to charge exchange scattering into
the charged pion channels.  The vast majority of the pions comes from initial $\Delta$
excitation (dash-dotted line), their production in the rescattering of nucleons is not
significant at these energies.

\begin{figure}[tb]
  \includegraphics[width=\columnwidth]{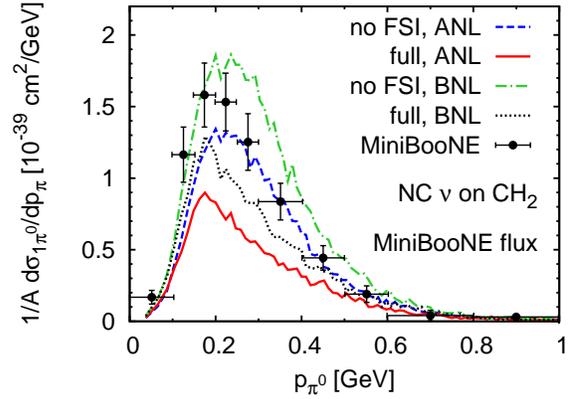}
  \caption{NC induced single-$\pi^0$ production on CH$_2$ as a function of the pion
    momentum averaged over the MiniBooNE flux. The dashed and the solid lines denote the
    calculation with a modified dipole form factor (fitted to ANL data) for the $\Delta$
    resonance (identical to the result in the left panel of
    \reffig{fig:MiniBooNE_K2K_NC}). The dash-dotted and dotted lines are obtained with a
    dipole form for the form factor (fitted to the BNL data).  The MiniBooNE data are
    taken from \refcite{andersonNUINT09Talk}.
  \label{fig:MiniBooNE_NC_ff}}
\end{figure}

The right panel of \reffig{fig:MiniBooNE_K2K_NC} shows the results for NC single-$\pi^0$
production off \oxygen{} averaged over the K2K energy flux which peaks at about 1.2 GeV
neutrino energy \cite{Nakayama:2004dp}. Compared to the left panel, the spectrum is
broader and extends to larger $T_\pi$ due to the higher neutrino energy. Again, pion
production through initial QE scattering is not sizable.

In \reffig{fig:MiniBooNE_NC_ff} we show also the latest NC$\pi^0$ differential cross sections from MiniBooNE
\cite{andersonNUINT09Talk}. It can be seen that the full line, depicting the results of our
calculation after FSI with the fit of the elementary cross section to the ANL data lies
nearly a factor 2 below the data. The dashed line gives the result of the same
calculation before FSI, the dash-dotted line shows the same quantity, but using a fit
to the BNL data for the elementary cross section. We find it remarkable that these two
curves (without FSI) are comparable to the data (with FSI), even though pion FSI have a
major effect on the spectra (see  \reffig{fig:MiniBooNE_K2K_NC}). The absolute height of the experimental cross
section is thus hard to reconcile with present descriptions of the elementary cross
section and with what we know from other reactions about the pion FSI.


\section{Duality in lepton scattering on the nucleon}

At higher energies that correspond to invariant masses of the $W,Z\, N$ system $>2\GeV$
the DIS channel becomes essential. A connection between the asymptotic DIS region and
the resonance--dominated mass range is provided by the so-called Bloom-Gilman quark-hadron duality.

Nearly forty years ago, Bloom and Gilman found \cite{Bloom:1970xb} that in electron
scattering on protons the inclusive structure function $F_2$ in the
resonance region oscillates around the DIS scaling curve and, after
averaging, closely resembles it. Furthermore, the resonance region data ``slide'' along the DIS curve with increasing $Q^2$.
If duality is understood quantitatively, there may be various applications.
For example, the region of high Bjorken variable $x$ is hardly experimentally investigated,
because in the DIS region it would require very high $Q^2$ and thus huge luminosities.
If duality is satisfied with good accuracy, one would be able to use the data in the resonance
region  to reach high $x$ at reasonable $Q^2$.

The topic becomes even more interesting when turning to nuclear targets and neutrino
sources.  The current precision measurements of the oscillation
parameters require an efficient and accurate description of the
neutrino--nucleus cross sections. Of particular interest is the
resonance region and the possibility of linking it with the DIS
region. A hadronic description of a neutrino-nucleus cross section
at low $Q^2$ requires a good knowledge of vector and axial transition form factors for
each resonance. For the majority of the resonances, these transition
form factors are not well constrained. Provided that one can establish
that quark-hadron duality holds with a reasonable accuracy, one could
think of using the DIS results for estimating the neutrino-nucleus
cross sections in the  transition region.

So far, most theoretical studies of quark--hadron duality in lepton  scattering were dealing with
nucleon targets~\cite{Matsui:2005ns,Graczyk:2005uv,Graczyk:2009px,Lalakulich:2006yn}. The DIS parts were considered
as known, the structure functions in the scaling region being conventionally
 evaluated from leading twist (LT) parton distribution functions (PDF): for example,
$F_2^{eN(LT)} = ( F_2^{ep} + F_2^{en})/2=5x/18\cdot(u + \bar{u} + d + \bar{d} + 2s/5 + 2\bar{s}/5)$,
$F_2^{\nu N(LT)} = ( F_2^{\nu p} + F_2^{\nu n})/2 = x (u + \bar{u} + d + \bar{d} + s + \bar{s})$.
For nucleons, several parametrizations of the PDFs are generally available (from the GRV, CTEQ and MRST groups).
In the region of moderate $x$, which is of interest for our duality study, they provide nearly the same results.

The studies of the resonance region differ in the way the models treat the resonant contributions and the way
they extract the structure functions.

The fact, that the resonance data for increasing $Q2$ "slide" along the DIS curve, can be
observed if a few~\cite{Lalakulich:2006yn} or even only one~\cite{Matsui:2005ns} resonance
are taken into account. The advantage of the model~\cite{Lalakulich:2006yn} is that the
structure functions are given as simple analytical functions of the momentum transfer
squared $Q^2$ and the energy transfer $\nu$, provided that the resonance form factors are
known. In this work the first four resonances were considered.  Generally, however, as it
was argued by Close \cite{Close:2003wz}, inclusion of several resonances of different
parities is desirable.

A quark model for resonance excitation has been applied by
\cite{Graczyk:2005uv,Graczyk:2009px} for the investigation of the duality in electro- and
neutrinoproduction.

Within the GiBUU framework, as it was mentioned above, 13 resonances can be considered for both electron and neutrino reactions.
The cross section is calculated numerically and the structure function $F_2$ is extracted from
the cross section in a convenient way:

\begin{equation}
\begin{array}{c}
\displaystyle
      \frac{d\sigma^N}{dQ^2 d\nu} = k_{EM,CC}  \frac{\pi}{E E'}\, \frac{F_2(Q^2,x)}{\nu}
 	\\[4mm]
\displaystyle
 	\left[  1-\frac{Q^2}{4EE'}+2\frac{Q^2}{4EE'}\frac{Q^2 \nu^2 + Q^4}{Q^4 (1+R)} \right]
\end{array}
\label{F2-from-xsec}
\end{equation}

The ratio $R$, defined as $2xF_1 (1+R) = F_2 (1+4m_N^2 x^2/Q^2)$, is the world average value
\[
\begin{array}{r}
\displaystyle
R (Q^2, x)= \frac{0.0635}{\ln (Q^2/0.04)} \left[ 1. + \frac{12. Q^2}{Q^2+1.0}\frac{0.125^2}{0.125^2 + x^2}\right]
 \\[4mm]
\displaystyle
 + \frac{0.5747}{Q^2} - \frac{0.3534}{Q^4+0.09},
\end{array}
\]
taken from \cite{Whitlow:1991uw}. The coefficients
\[
 k_{EM}=\frac{4\alpha_{em}^2 E'{}^2}{Q^4} , \qquad k_{CC}=\frac{G_F^2 E'{}^2}{2\pi^2}  ,
\]
are the Mott cross sections for electron and neutrino reactions. For electroproduction on
an isoscalar target, $\sigma^N=(\sigma^{ep}+\sigma^{en})/2$ is half sum of
electroproduction cross sections on proton and neutron.  For charged current
neutrinoproduction, in order to eliminate the structure function $F_3$, one should use the
linear combination of the neutrino and anti-neutrino cross sections. For an isoscalar
target it is sufficient to take $\sigma^N=(\sigma^{\nu p} + \sigma^{\bar{\nu} p} +
\sigma^{\nu n} + \sigma^{\bar{\nu} n} )/4 $.

For a quantitative estimate of the validity of duality it is convenient to introduce the ratio
of the integrals of the resonance (res) and DIS structure functions
\begin{equation}
I_i(Q^2) =
\frac{ \int_{\xi_{\rm min}}^{\xi_{\rm max}} d\xi\
    {\cal F}_i^{(\rm res)}(\xi,Q^2) }
     { \int_{\xi_{\rm min}}^{\xi_{\rm max}} d\xi\
    {\cal F}_i^{(\rm DIS)}(\xi,Q^2_{DIS}) }\ ,
\label{eq:Int}
\end{equation}
where ${\cal F}_i$ denotes $2xF_1$, $F_2$ or $x F_3$ (for neutrino scattering). The value $Q^2_{DIS}$ is
taken as the actual $Q^2$ value for a given parametrization of DIS PDFs
(for nucleon, in our case, $Q^2_{DIS}=10 \GeV$) or a DIS experimental data set (for nuclei).
Under conditions of perfect quark--hadron duality this ratio would be $1$
and independent of $Q^2$. Thus, the degree to which the local duality
is fulfilled can be estimated from the $Q^2$ dependence and the deviation
from $1$ of the computed $I_2$.

\begin{figure}[htb]
  \includegraphics[angle=-90,width=\columnwidth]{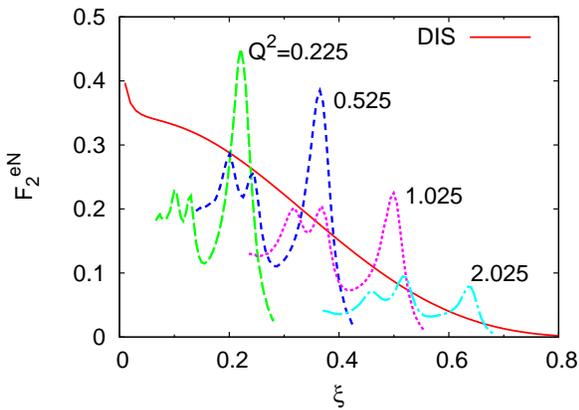}
\caption{ $F_2^{eN}$ as a function of $\xi$, for $Q^2 =$ 0.225, 0.525, 1.025
    and $2.025\GeV^2$ (indicated on the spectra), compared with the leading twist
    parametrizations at $Q^2=10\GeV^2$.
}
\label{fig:nucleon-em}
\end{figure}

The isoscalar nucleon $F_2^{eN}$ structure function, which includes both resonance and background contributions,
is shown in Fig.~\ref{fig:nucleon-em} versus the Nachtmann variable $\xi$.
Notice, that $\xi$ decreases with increasing invariant mass $W$. For a given $Q^2$ value, the highest peak at the larger $\xi$
value corresponds to the $\Delta-$resonance peak, and the two lower peaks at smaller values of $\xi$ correspond to the second
($1.40\GeV \lesssim W \lesssim 1.56 \GeV$)
and the third ($1.56 \GeV \lesssim W \lesssim 2.0 \GeV$) resonance regions.
The general picture shows a reasonable agreement with the duality hypothesis.

\begin{figure}[htb]
  \includegraphics[angle=-90,width=\columnwidth]{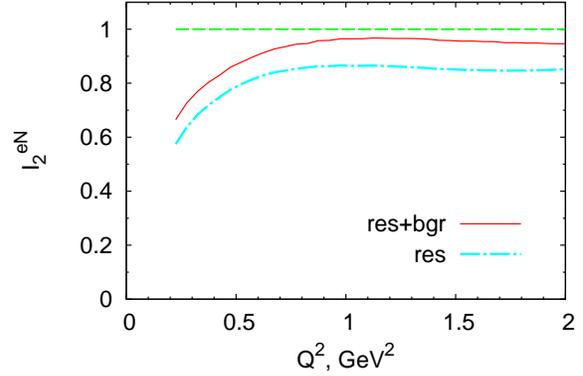}
\caption{
    Ratio $I_2^{eN}$ of the integrated $F_2^{eN}$ in the resonance
    region to the leading twist functions.
}
\label{fig:I-nucleon-em}
\end{figure}

In Fig.~\ref{fig:I-nucleon-em}, the ratio of the integrals $I_2^{eN}$, defined in
(\ref{eq:Int}), is shown not only for the whole structure function (resonance $+$ 1-pion
background), but also for the resonance contribution separately.  For $Q^2>0.5 \GeV^2$,
the ratio $I_2^{eN}$ for the resonance contribution only is at the level of $0.85$, which
is smaller and flatter in $Q^2$ in comparison with the results \cite{Lalakulich:2006yn,
  Lalakulich:2007zz} of the Dortmund group resonance model.  The difference is due to the
different parametrization of the electromagnetic resonance form factors used in the two
models.  The $1 \pi$ background gives a noticeable contribution and brings the ratio up to $0.95$.
The fact, that the ratio is smaller than $1$ is no surprise, because additional
non-resonant contributions like 2- and many-pion background are possible, but not taken
into account here.  They are the subject of coming investigations. Indeed, they could be
obtained from requirement $I_2=1$.

The principal feature of neutrino reactions, stemming  from fundamental isospin arguments,
is that duality does not hold for proton and neutron targets separately.
The interplay between the resonances of different isospins allows for duality to
hold with reasonable accuracy for the average over the proton and
neutron targets. We expect, that a similar
picture emerges in neutrino reactions with nuclei.

\begin{figure}[htb]
\includegraphics[angle=-90,width=\columnwidth]{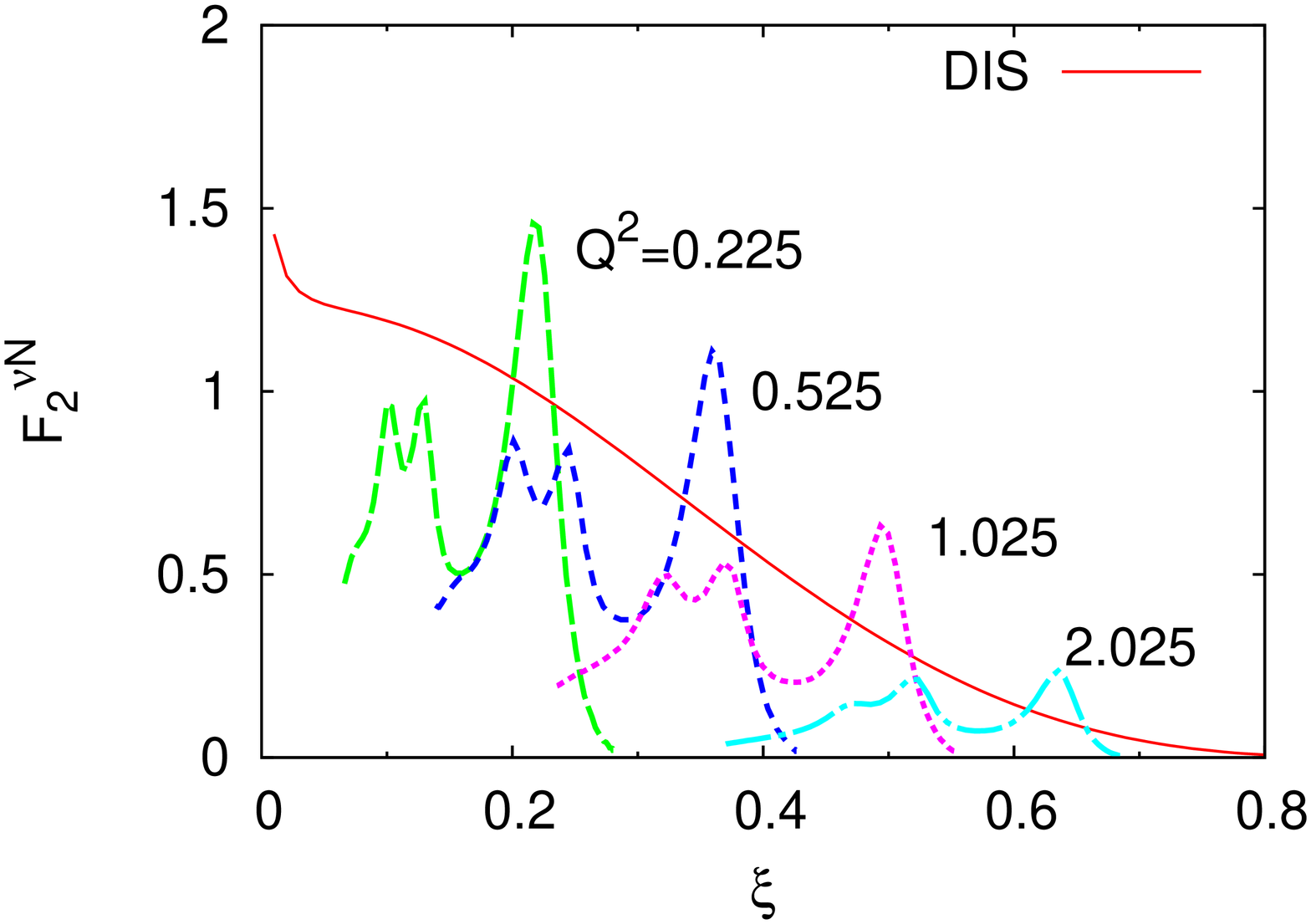}
\label{fig:nucleon-weak}
\caption{$F_2^{\nu N}$ as a function of $\xi$, for $Q^2 =$ 0.225, 0.525, 1.025
    and $2.025\GeV^2$ (indicated on the spectra), compared with the leading twist
    parametrizations at $Q^2=10\GeV^2$.
}
\end{figure}

For neutrinoproduction, the structure function $F_2^{\nu N}$ is shown in
Fig.~\ref{fig:nucleon-weak} for the resonance contribution only.

\begin{figure}[htb]
 \includegraphics[angle=-90,width=\columnwidth]{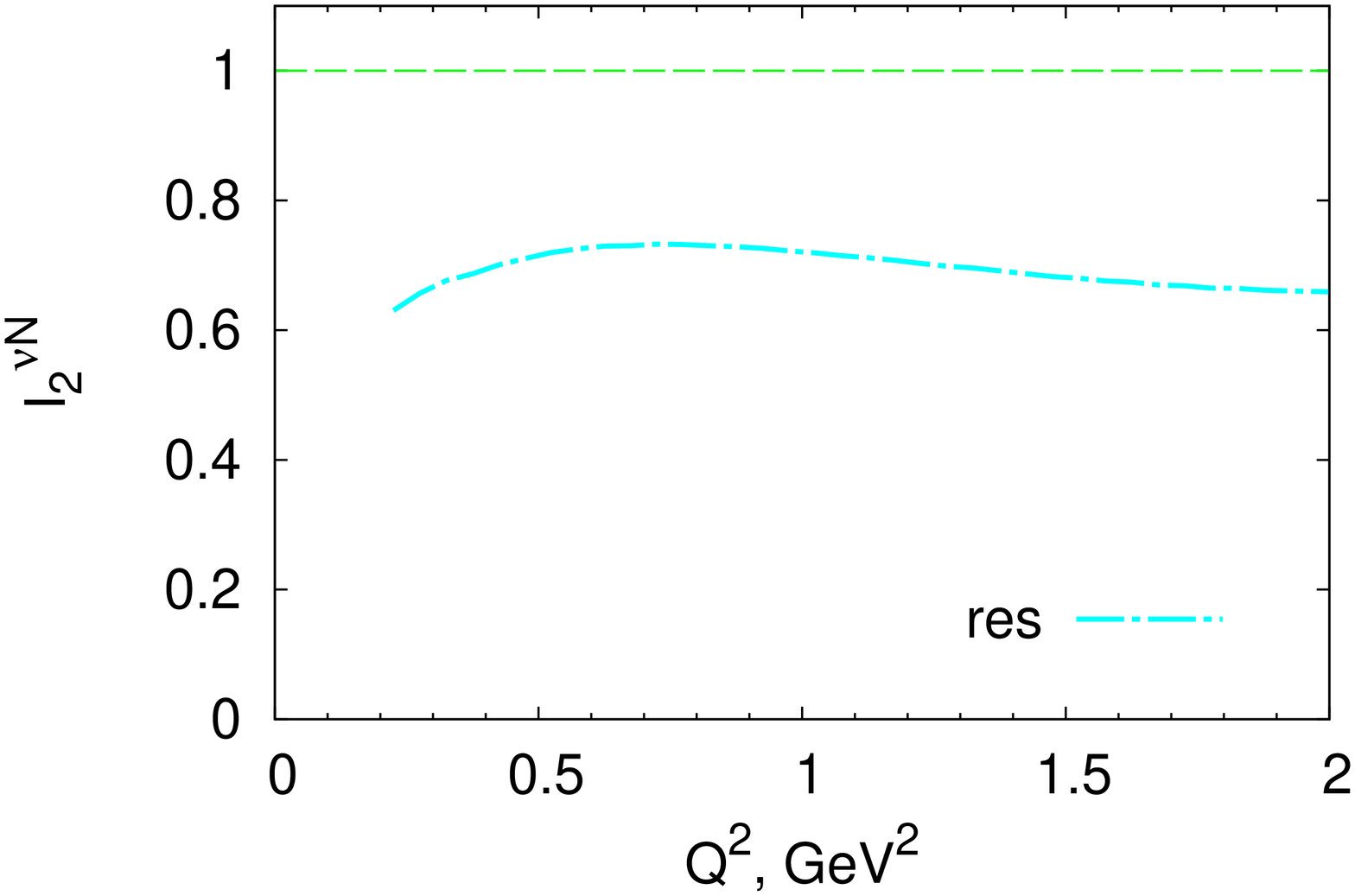}
\caption{Ratio $I_2^{\nu N}$ of the integrated $F_2^{\nu N}$ in the resonance
    region to the leading twist functions.
}
\label{fig:I-nucleon-weak}
\end{figure}

The ratio $I_2^{\nu N}$ is shown in Fig.~\ref{fig:I-nucleon-weak}
and appears to be at the level  of $0.7$, which is (similar to the electron case) smaller than $0.8$, which has been calculated within the Dortmund
resonance model \cite{Lalakulich:2006yn, Lalakulich:2007zz}. Thus, one would expect a large contribution
from the background. The role of the background in the anti-neutrino channel is under investigation.

\section{Duality in lepton scattering on the nucleus}

Recent electron scattering measurements at JLab have confirmed the validity of the
Bloom--Gilman duality for proton, deuterium and iron structure functions. Further
experimental efforts are required for neutrino scattering. Among the upcoming neutrino
experiments, Miner$\nu$a\cite{minerva} and SciBooNE\cite{sciboone} aim at measurements
with carbon, iron and lead nuclei as targets.

One of the major issues for nuclear targets is the definition of the nuclear structure functions $F_{1(2,3)}^A$.
Experimentally they are determined from the corresponding cross sections, using Eq.~(\ref{F2-from-xsec}).

We follow the same procedure, using the GiBUU cross sections.
So, at the first step the inclusive double differential cross section $d\sigma / dQ^2 d\nu$ is calculated within
the GiBUU model. The FSI do not play any role for the inclusive cross section. The in-medium corrections are
of considerable importance.

\begin{figure}[hbt]
	\includegraphics[angle=-90,width=\columnwidth]{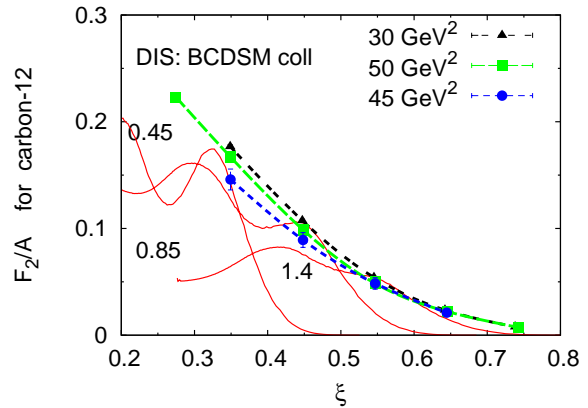}
        \caption{Resonance curves $F_2^{e {}^{12}C}/12$ as a function of
          $\xi$, for $Q^2 =$ 0.45, 0.85 and $1.4\GeV^2$ (indicated on the
          spectra), compared with the experimental BCDMS data
          \cite{Bollini:1981cr,Benvenuti:1987zj} in the DIS region at $Q^2_{DIS}=30$, $45$
          and $50\GeV^2$. }
\label{fig:F2-C12-em}
\end{figure}

In Fig.~\ref{fig:F2-C12-em}, the resonance contribution to the
$F_2^A/A$ structure functions  for a carbon target  are shown for several $Q^2$ values. They are compared to experimental
data obtained by the BCDMS collaboration \cite{Bollini:1981cr,Benvenuti:1987zj} in muon--carbon
scattering in the DIS region ($Q^2\sim 30-50\GeV^2$). They are shown as experimental points connected by
smooth curves. For different $Q^2$ values, the experimental curves agree within
$5\%$ in most of the $\xi$ region, as expected from Bjorken scaling.

When investigating duality for a free nucleon, we took the average over free proton and
neutron targets, thus considering the isoscalar structure
function. Since the carbon nucleus contains an equal number of protons
and neutrons, averaging over isospin is performed automatically.
Due to the Fermi motion of the target nucleons, the peaks from the various
resonance regions, which were clearly seen for the nucleon target, are hardly distinguishable
for the carbon nucleus.

As expected  from local duality, the resonance structure functions for the various
$Q^2$ values slide along a curve, whose $\xi$ dependence is very similar to the scaling--limit
DIS curve. However, for all $\xi$, the resonance curves lie below the
experimental DIS data.


To quantify this underestimation,  we now consider the ratio of the integrals of
the resonance (res) and DIS structure functions, determined in Eq.~(\ref{eq:Int}).
As it is explained in \cite{Lalakulich:2008tu}, the integration limits are to be determined
in terms of the effective $\tilde{W}$ variable, experimentally (see, for example, \cite{Sealock:1989nx})
defined as  $\tilde{W}^2=m_N^2+2m_N\nu-Q^2$. For a free nucleon $\tilde{W}$  coincides with the
invariant mass $W$. For a nucleus, it differs from $W$ due to the Fermi motion of bound nucleons,
but still gives a reasonable
estimation for the invariant mass region involved in the problem.

In particular, the resonance curves presented in all figures are plotted in the
region from the pion--production threshold up to $\tilde{W}=2\GeV$.
For a free nucleon, the threshold value for 1-pion production (and thus
the threshold value of the resonance region) is $\tilde{W}_{\rm min}=W_{\rm min} \approx 1.1 \GeV$.
Bound backward-moving nucleons in a nucleus allow  lower $W$ values beyond the free--nucleon limits.
The threshold for the structure functions is now defined in terms of $\nu$
or $\tilde{W}$,  rather than $W$.
Hence, we consider two different cases in choosing the
$\xi$ integration limits for the ratio (\ref{eq:Int}). First, for a
given $Q^2$, we choose the $\xi$ limits in the same manner as for a free nucleon:
\begin{equation}
\begin{array}{l}
\displaystyle
\xi_{\rm min} = \xi(\tilde{W}=1.6\GeV, \, Q^2), \qquad
\\[3mm]
\displaystyle
\xi_{\rm max} = \xi(\tilde{W}=1.1\GeV, \, Q^2) \ .
\end{array}
\label{11-16-}
\end{equation}
We refer to this choice as integrating ``from 1.1 GeV''. The integration
limits for the DIS curve always correspond to this choice.  As a
second choice, for
each $Q^2$ we integrate the resonance curve from the threshold, that
is from as low $\tilde{W}$ as is achievable for the nucleus under
consideration. This corresponds to the threshold value at higher $\xi$
and is referred to as integrating ``from threshold''. With this choice we
guarantee that the extended kinematical regions typical for resonance
production from nuclei are taken into account.
Since there is no natural threshold for the $\xi_{min}$, for
both choices it is determined from $\tilde{W}=1.6\GeV$, as defined in
Eq.~(\ref{11-16-}).

\begin{figure}[h!bt]
	\includegraphics[angle=-90,width=\columnwidth]{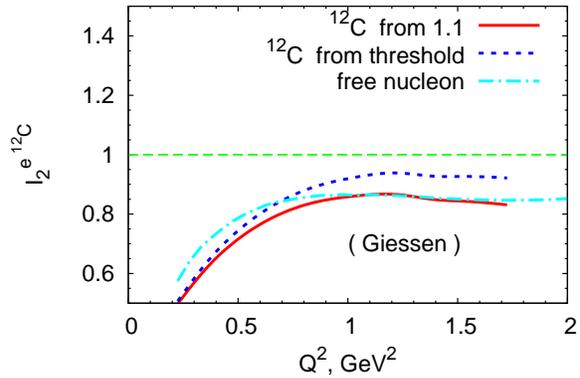}
        \caption{Ratio
 defined in Eq.(\ref{eq:Int}) for the free nucleon (dash-dotted line),
 and $^{12}C$.  We consider the lower limits determined by
 $\tilde{W}=1.1 \GeV$ (solid line) and by the threshold value (dotted
 line).}
\label{fig:F2-C12-int}
\end{figure}

The results for the ratio (\ref{eq:Int}) are shown in Fig.~\ref{fig:F2-C12-int}.
The curve for the isoscalar
free-nucleon case is the same as in Ref.~\cite{Lalakulich:2006yn} with
the ``GRV'' parametrization for the DIS structure function. One can
see that the carbon curve obtained by integrating ``from threshold''
lies above the one obtained by integrating ``from 1.1 GeV'', the
difference increasing with $Q^2$. This indicates that the threshold
region becomes more and more significant, as one can see from
Fig.~\ref{fig:F2-C12-int}.
Recall, that the flatter the curve is and the closer it
gets to 1, the better local duality would hold.
Our calculations for carbon show that the ratio is at the same level as that for the free nucleon or even higher.

\begin{figure}[htb]
        \includegraphics[angle=-90,width=\columnwidth]{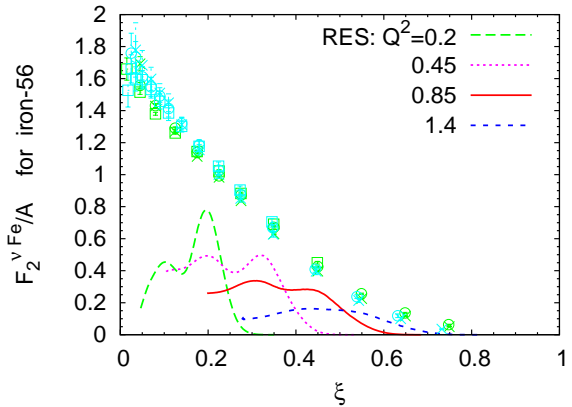}
\caption{$F_2^{\nu \, {}^{56}\! Fe}/56$  as a function of $\xi$
 	for $Q^2 =$ 0.2, 0.45, 0.85, and $1.4\GeV^2$. The
 	calculations are compared with the DIS data from CCFR and NuTeV experiments
 	Refs.~\cite{Seligman:1997mc,Tzanov:2005kr}. The DIS data refer to
 	measurements at $Q^2_{DIS}=7.94$, $12.6$ and $19.95 \GeV^2$.
}
\label{fig:Fe56-nu}
\end{figure}

For neutrino--iron scattering, the structure functions $F_2^{\nu Fe}$  are shown in Fig.~\ref{fig:Fe56-nu}.
As for the electron-carbon results of  Fig.~\ref{fig:F2-C12-em}, the resonance structure is hardly
visible.  The resonance structure functions are compared to the experimental data in the DIS region obtained by the CCFR
\cite{Seligman:1997mc} and NuTeV \cite{Tzanov:2005kr} collaborations.
It appears, that the resonance curves slide along the DIS curve, as one would
expect from local duality, but lie well below the DIS measurements. Hence, the
computed structure functions do not average to the DIS curve. The necessary condition for
local duality to hold is thus not fulfilled.

As expected  from local duality, the resonance structure functions for the various
$Q^2$ values slide along a curve, whose $\xi$ dependence is very similar to the scaling--limit
DIS curve. However, for all $\xi$, the resonance curves lie below the
experimental DIS data.

\begin{figure}[htb]
        \includegraphics[angle=-90,width=\columnwidth]{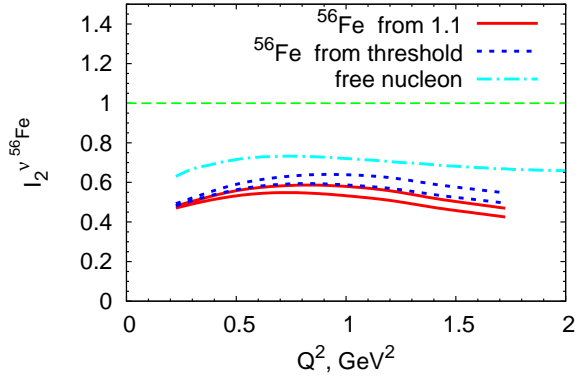}
\caption{Ratio $I_2^{\nu \, {}^{56}Fe}$ defined in Eq.~(\ref{eq:Int}) for the free nucleon
	(dash-dotted line) and $^{56}$Fe.  The results are
    	displayed for two choices of the lower limit in the integral:
    	$\tilde{W}=1.1 \GeV$ (solid line) and threshold (dotted line).  For
    	each of these two choices we have used two sets of DIS data in
    	determining the denominator of Eq.~(\ref{eq:Int}).  These sets of
    	DIS data are obtained at $Q^2_{DIS}=12.59$ and 19.95 GeV$^2$.
}
\label{fig:I-Fe56-nu}
\end{figure}

The ratio $I_2^{\nu \, {}^{56}\! Fe}$  defined in Eq.(\ref{eq:Int}) is shown in Fig.~\ref{fig:I-Fe56-nu}.
The curve for the isoscalar free nucleon case is also presented for comparison.
Our results show, that 1) this ratio is
significantly smaller than 1 for all $Q^2$; 2) it is significantly smaller than
the one for the free nucleon; 3) $I_2$ is even lower than the
corresponding ratio for electroproduction; 4) $I_2$
slightly decreases with $Q^2$.

To summarize, we find  that the resonance structure functions are consistently smaller than DIS
functions in the same region of  the Nachtmann variable $\xi$.
This is in agreement with earlier work \cite{Lalakulich:2008tu}, which implements elementary
resonance vertices and nuclear effects differently.
Recall that in this analysis for nuclei we include the resonance structure functions
and ignore the background ones. To estimate their contribution and compare the results with
the nucleon case is one of the primary tasks of coming investigations.

\section{Conclusions}

Overall, the impact of nuclear effects impact on observables is dramatic. Final state interactions of produced pions lead to pions being absorbed in the nucleus or rescattered, thus
reducing the cross section and shifting the observed pion distributions
to lower values of the pion energy. Furthermore, such interactions can lead to knock-out nucleons which experiments tend to identify with QE events. CCQE and pion production are thus closely entangled. Thus model-independent data are
needed for a meaningful comparison of theory with current experiments.
Initial state interactions reduce the cross sections and structure functions in the resonance
region and change the form of the distributions, dissolving the peaks of individual resonances. The Bloom-Gilman duality seems to be violated
for nuclear targets, possibly indicating  a major change of background amplitudes in the medium.


\begin{theacknowledgments}
  This work has been supported by the Deutsche Forschungsgemeinschaft.
\end{theacknowledgments}

\end{document}